# *A very, very peculiar telescope of the 1610s*

Paolo Del Santo

Museo Galileo – Istituto e Museo di Storia della Scienza (Florence - Italy)

**Abstract**

A re-examination of a well-known iconographic source, the *Allegory of Sight* by Jan Brueghel the Elder and Peter Paul Rubens, reveals that one of the two telescopes depicted in the painting has a highly unusual — and until now unknown to the historians of the telescope— particularity.

In the painting the *Allegory of Sight*, depicted in Antwerp in 1617 by Jan Brueghel the Elder (1568–1625) and Peter Paul Rubens (1577–1640), two telescopes are clearly visible in the foreground (Figure 1). Because of the extremely small number of surviving telescopes from 1610s, the painting is an important and faithful iconographic source for historians of the telescope.

The *Sight* is part of the series of allegoric paintings *The Five Senses*, now at the Museo Nacional del Prado of Madrid (Spain), which is one of the most famous collaborations by Brueghel and Rubens, who were closed friend, and executed about two dozen of paintings together during over than a quarter of a century, from about 1598 to 1625.

Following a tried and tested technique, Brueghel, renowned for his exceptionally meticulous attention to details, began the work creating the settings, and, in a later stage, Rubens painted the figures; finally, Brueghel applied the conclusive brushstrokes to integrate the figures depicted by Rubens into areas of the painting surrounding them.

The *Sight* is set in a *Kunstkammer*, i.e. a cabinet of curiosities, filled with numerous precious objects that recall the splendour of the court of the sovereigns of Habsburg Netherlands, the Archduke of Austria Albert VII and his wife Isabella, Infanta of Spain and Portugal: paintings, ancient marble busts, spectacles, armillary spheres and other scientific instruments, etc.

As can be seen, the two telescopes portrayed in the painting are very different from each other in size and characteristics. The one on the left is a seven draw tubes, of exquisite workmanship, with a seemingly metallic tube, maybe even silver, lens housing in turned ebony, and an elegant pedestal. In short, a very sophisticated object, clearly intended for rich and refined customers.

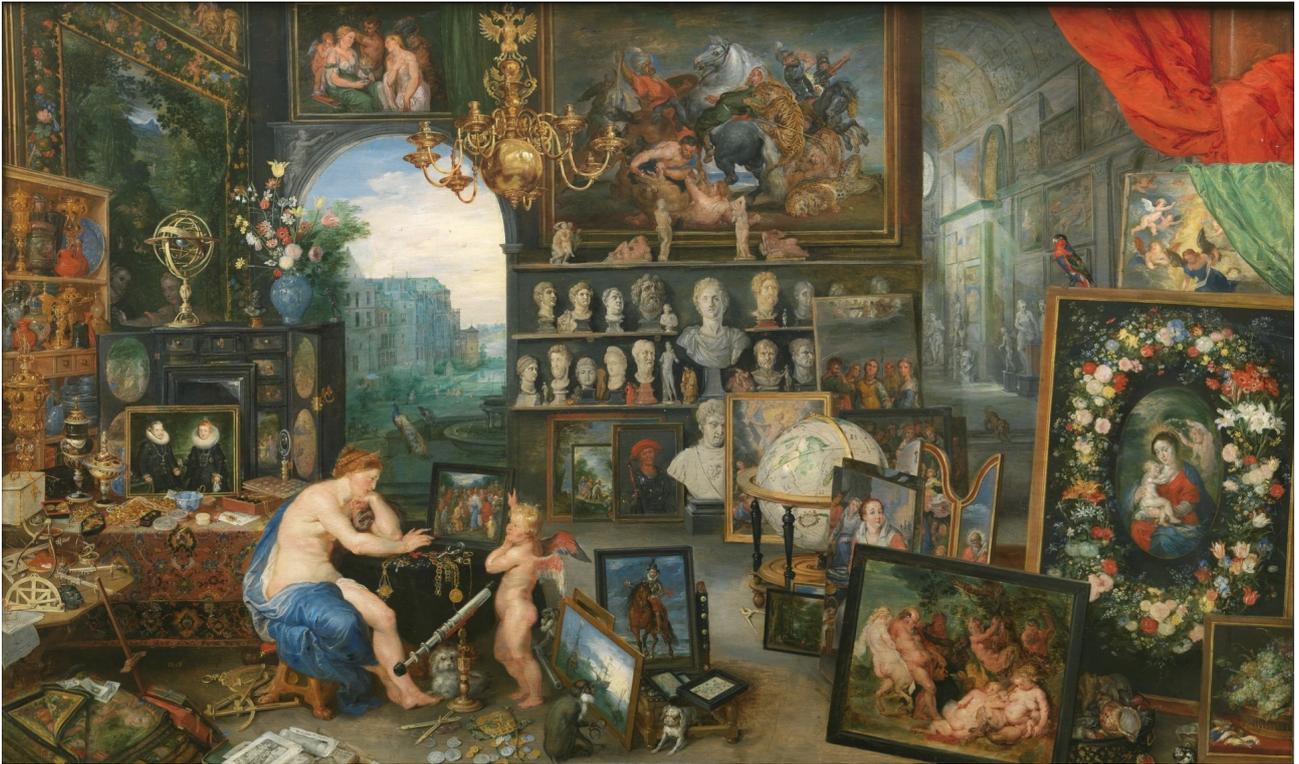

Figure 1: Brueghel J. the Elder and Rubens P. P. ( 1617). *Allegory of Sight.* [Oil on panel, 109.5 cm × 64.7 cm], Museo Nacional del Prado, Madrid (Spain).

(https://www.museodelprado.es/en/the-collection/art-work/the-sense-of-sight/494fd4d5-16d2-4857-811b-e0b2a0eb7fc7).

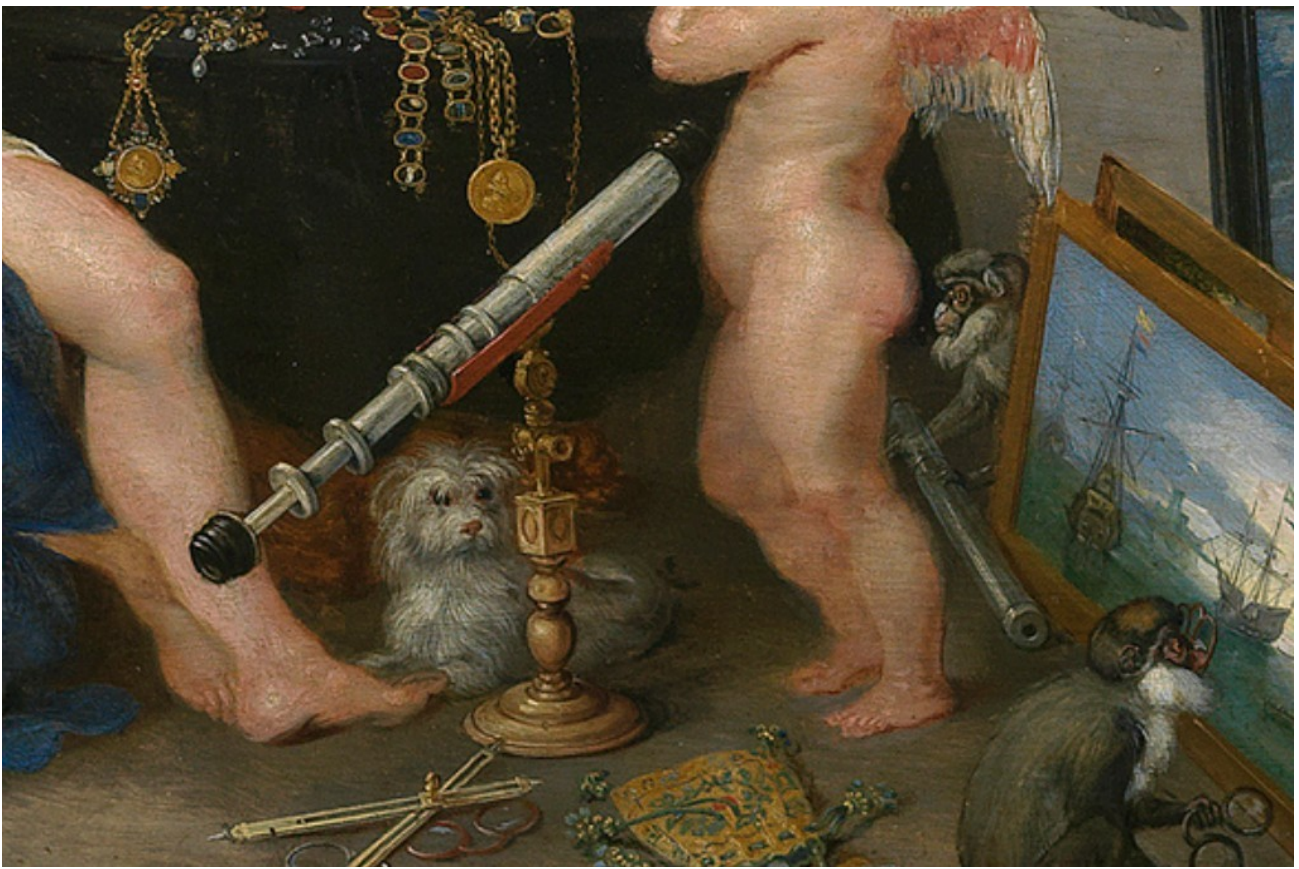

Figure 2: Jan Brueghel the Elder and Peter Paul Rubens, *Allegory of Sight*, detail.

The other telescope, hold by a monkey, is smaller, made up of only two elements, the main tube and the ocular one, and without mount. At first glance, it appears spartan, totally devoid of any ornament or decoration. In other words, a modest instrument which seems to contrast with refinement of the other telescope and the richness of the scene.

Both the instrument were examined by Silvio Bedini (1971: 170) over fifty years ago, and, more recently, by Paolo Molaro and Pierluigi Selvelli in a series of papers, published between 2009 and 2011. However, Bedini made the gross error of mistaking the telescope held by the monkey for a microscope (but it is definitely too big to be such an instrument), while Molaro and Selvelli interpret the image correctly as the representation of a telescope, nevertheless understanding their opinion about it is far from easy. I will try here to summarize the evolution —or at least the multiplicity— of their judgments (in fact, almost all Molaro and Selvelli's writings on the subject were published in proceedings of symposiums, which, as is known, are often printed long after the conference date. For this reason, the one here outlined is just a possible, plausible chronology of the evolution of their positions).

In the proceedings of a symposium held in Paris in early 2009, Molaro and Selvelli (2011b: 330) claimed that

> [a] close inspection on a high quality reproduction of the painting has [...] clearly shown that the tube appears to be made of metal, possibly tin [...]. The object is clearly a spyglass and its aspect and size clearly indicate that, most likely, it corresponds to the spyglass depicted in [the *Extensive Landscape with View of the Castle of Mariemont*].

The *Extensive Landscape with View of the Castle of Mariemont* is a painting by Jan Brueghel the Elder (not in partnership with Rubens), datable between 1609 and 1611, and the "spyglass" to which Molaro and Selvelli refer is the small handheld telescope through which the Archduke Albert looks in the distance toward a wood (fig. 3).

In the proceedings of a symposium held in Venice (Italy), in October 2009, they argue less assertively that "[i]t is remarkable that it belongs to the Archduke's collection and it could be the same spyglass [depicted in the *Extensive Landscape with View of the Castle of Mariemont*] several years before" (Molaro & Selvelli, 2010: 249); which is essentially the same position expressed in a paper, in Italian, published in the following December ("the 'tube' held by the monkey is the same instrument or one very similar to that which the Archduke used in the [*Extensive Landscape with View of the Castle of Mariemont*]", Selvelli & Molaro, 2009: 29).

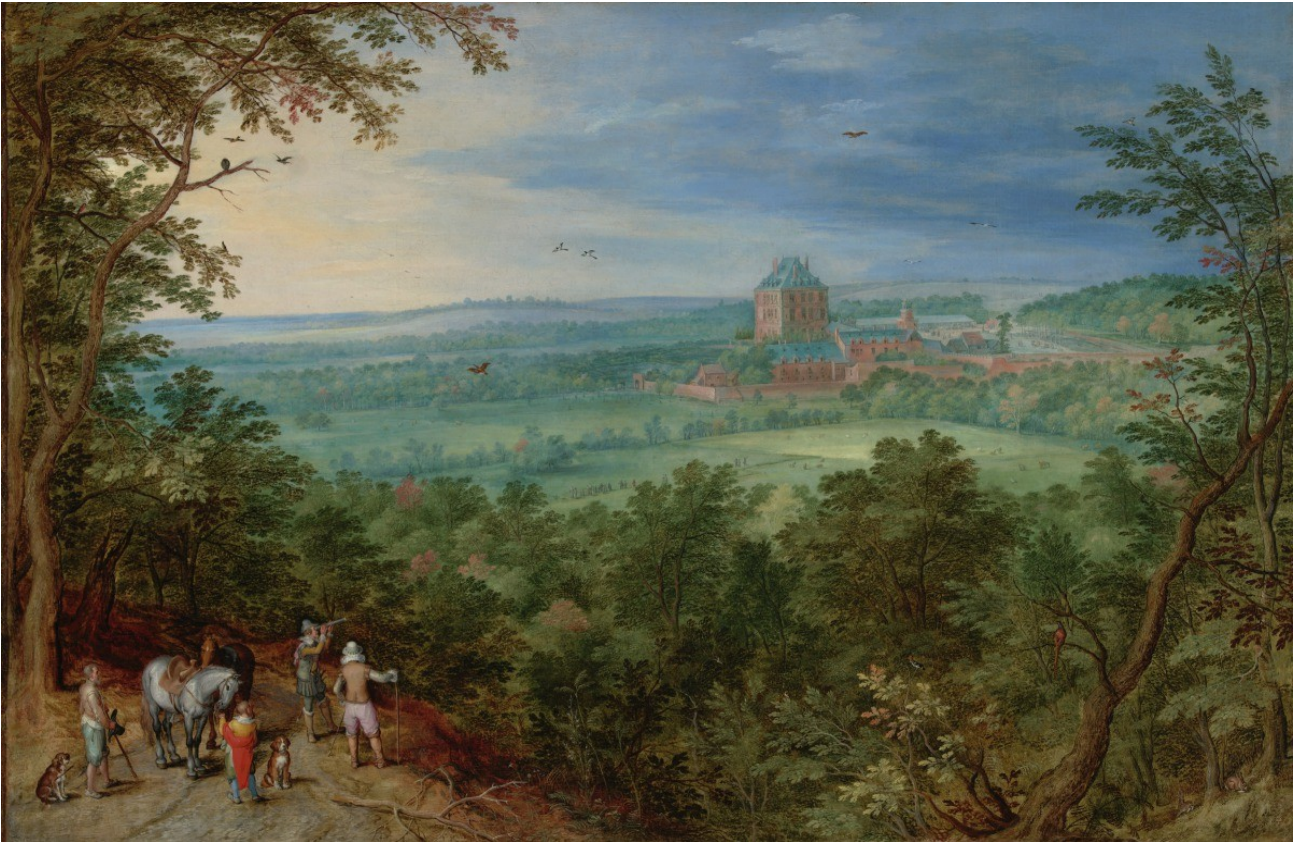

Figure 3: Brueghel J. the Elder, (1609-1611). *Extensive Landscape with View of the Castle of Mariemont*, [oil on canvas, 85 x 131 cm]. Virginia Museum of Fine Arts, Richmond, VA, United States. (https://vmfa.museum/piction/6027262-8052649/)

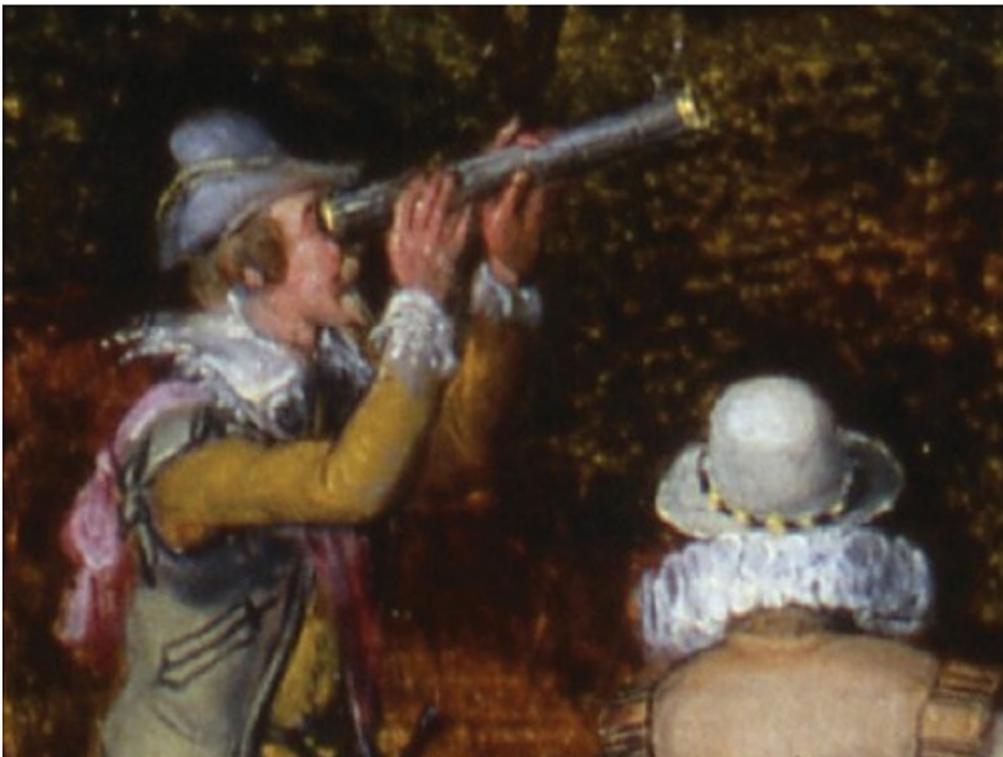

Figure 4: Brueghel J. the Elder, (1609-1611). *Extensive Landscape with View of the Castle of Mariemont*. Detail.

In actual fact, they are quite different at a glance, and it is difficult to even understand how Molaro and Selvelli could have initially believed that they were the same instrument. For instance, the lens cells of the telescope depicted in the *Extensive Landscape with View of the Castle of Mariemont* (although, on the canvas, it is long only about 4 cm, Brueghel's extraordinary ability at rendering the minutest details, allows us to discern its salient characteristics) are made of brass, of which there does not seem to be any trace in the telescope of the *Allegory* (fig. 4). Also their shape are quite different: the objective lens cell of the former clearly ends in a ferrule, while the one of the latter is perfectly cylindrical and of same diameter of the main tube.

In subsequent (?) papers, the two authors seem to become aware of the profound difference in appearance between the two instruments:

> "[i]t is rather strange that such a cheap instrument was reproduced beside a silver telescope [i.e. the above-mentioned seven draw tubes telescope] which holds a central position in the painting [...]. However, its presence in the painting implies it was of great importance to the Archduke who kept it in his collection of scientific instruments. It could be that this spyglass was the very same spyglass depicted several years before in the *Extensive Landscape with View of the Castle of Mariemont* since it has a similar size. However, its much poorer appearance and rather cheaper enclosure suggest it could be a second very early telescope". (Molaro & Selvelli, 2011a: 16)

In conclusion, whereas initially Molaro and Selvelli firmly maintained that the telescope depicted in the *Allegory* "most likely" was the very same instrument depicted in the *Extensive Landscape with View of the Castle of Mariemont*, gradually (but rapidly), they then turned towards the opposite view, namely that, due to the "much poorer appearance" of the former as compared to the latter, they are most probably two different instruments. Eventually, the two authors seems to have abandoned any hypothesis about the nature of this telescope and its possible or alleged resemblance with that one portrayed in the *Extensive Landscape* (Selvelli & Molaro, 2010).

Anyways, whatever is Molaro and Selvelli's conclusive view, we can notice that whenever they mention the 'monkey's telescope' they insist on the sameness or at least the similarity with the telescope depicted in the *Extensive Landscape with View of the Castle of Mariemont*, which, in their opinion, could even be the representation of "one of the first telescope ever made by mankind" (Selvelli & Molaro, 2011b)! However this instrument shows a peculiarity of which Molaro and Selvelli seem not to have realized: the wrist of the monkey that holds the telescope is clearly visible through the ocular tube. In other words, the latter was made of the only transparent material known at that time: glass!

Besides, on a closer examination, also the main tube appears to be made of glass, as it is evidenced by the fact that the corner of the frame of the maritime painting, rested vertically on the floor next to the telescope, is visible through it. That means that the whole instrument, except maybe the cells, is made of glass.

In conclusion, it is evident that, despite its apparent simplicity, this instrument was a curio, a rarity, fruit of the creative talent of a skill artificer; an instrument whose fragility made it a mere collector's item, of little or no practical use and intended just to be shown, destined to high ranking personage, as the Archduke Albert was. In other word, it was the rareness and sophistication of this instrument, and not its alleged cheapness —which, in Molaro and Selvelli's opinion, would be due to the fact it is one of the earliest telescope ever built, as we have seen— that justify its presence in his cabinet of curiosities.

Apparently, this is the only known example of the use of this material to make the tube of a telescope, but, now that we know that such an instrument existed, some glass tube lying forgotten in the warehouse of some museum, without anyone having ever understood the use to which it was intended, could be identified as the tube of a telescope in future.

2010). Available at: https://arxiv.org/ftp/arxiv/papers/1512/1512.01347.pdf (accessed 24 February 2024).